\documentclass[pre,showpacs,amsfonts,amssymb,eqsecnum,%
twocolumn,floatfix]{revtex4}
\pdfoutput=1
\usepackage{amsmath}
\usepackage{graphicx}
\usepackage{bm}
\usepackage{color}
\allowdisplaybreaks

\begin{document}

\title{Transport properties and first arrival statistics of random motion with stochastic reset times}
	
	\author{Axel Mas{\'o}-Puigdellosas}
	\affiliation{Grup de F{\'i}sica Estad\'{i}stica.  Departament de F{\'i}sica.
		Facultat de Ci\`encies. Edifici Cc. Universitat Aut\`{o}noma de Barcelona,
		08193 Bellaterra (Barcelona) Spain}
	\author{Daniel Campos}
	\affiliation{Grup de F{\'i}sica Estad\'{i}stica.  Departament de F{\'i}sica.
		Facultat de Ci\`encies. Edifici Cc. Universitat Aut\`{o}noma de Barcelona,
		08193 Bellaterra (Barcelona) Spain}
	\author{Vicen\c{c} M\'{e}ndez}
	\affiliation{Grup de F{\'i}sica Estad\'{i}stica.  Departament de F{\'i}sica.
		Facultat de Ci\`encies. Edifici Cc. Universitat Aut\`{o}noma de Barcelona,
		08193 Bellaterra (Barcelona) Spain}
	\date{\today}
	
       \begin{abstract}
        Stochastic resets have lately emerged as a mechanism able to generate finite equilibrium mean square displacement (MSD) when they are applied to diffusive motion. Furthermore, walkers with an infinite mean first arrival time (MFAT) to a given position $x$, may reach it in a finite time when they reset their position. In this work we study these emerging phenomena from a unified perspective. On one hand we study the existence of a finite equilibrium MSD when resets are applied to random motion with $\langle x^2(t)\rangle _m\sim t^p$ for $0<p\leq2$. For exponentially distributed reset times, a compact formula is derived for the equilibrium MSD of the overall process in terms of the mean reset time and the motion MSD. On the other hand, we also test the robustness of the finiteness of the MFAT for different motion dynamics which are subject to stochastic resets. Finally, we study a biased Brownian oscillator with resets with the general formulas derived in this work, finding its equilibrium first moment and MSD, and its MFAT to the minimum of the harmonic potential.
        \end{abstract}
        
        \maketitle
\section{Introduction}
The strategies employed by animals when they seek for food are complex and strongly dependent on the species. A better understanding of their fundamental aspects would be crucial to control some critical situations as the appearance of invading species in a certain region or to prevent weak species to extinct, for instance. 

In the last decades, a lot of effort has been put in the description of the territorial motion of animals \cite{OkLe02}. Among other, random walk models as correlated random walks and L\'evy walks \cite{BaDa05,MeCaBa14} or L\'evy flights \cite{ReRh09} are commonly used. Nevertheless, in the vast majority of these approaches, only the foraging stage of the territorial dynamics is described (i.e. the motion patterns while they are collecting), leaving aside the fact that some species return to their nest after reaching their target.

Having in mind that limitation, Evans and Majumdar \cite{EvMa11} studied the properties of a macroscopic model consisting on a diffusive process subject to resets with constant rate (mesoscopically equivalent to consider exponentially distributed reset times), which introduces this back-to-the-nest stage. For this process, the mean first passage time (MFPT) is finite and the mean square displacement (MSD) reaches an equilibrium value. The latter result allows us to define the home range of a given species being a quantitative measure of the region that animals occupy around its nest. 

From then on, multiple works have been published generalizing this seminal paper \cite{EvMa11p,WhEv13,EvMa13,GuMa14,EvMa14,
DuHe14,Pa15,MaSa15p,ChSc15,PaKu16,NaGu16,
FaEv17,BoEv17}, by introducing for instance absorbing states \cite{WhEv13} or generalising it to $d$-dimensional diffusion \cite{EvMa14}. Some works have also been devoted to the study of L\'evy flights when they are subject to constant rate resets \cite{KuMa14,KuGu15} and others have focused on the analysis of first passage processes subject to general resets \cite{RoRe15,Re16,PaRe17}. Also, stochastic resets have been studied as a new element within the continuous-time random walk (CTRW) formulation \cite{MoVi13,CaMe15,MeCa16,MoMa17,Sh17}.

Despite the amount of works devoted to this topic, the existence of an equilibrium MSD and the finiteness of the MFAT found in \cite{EvMa11} for diffusive processes with exponential resets have not been explicitly tested in general. In this work we address this issue by analyzing these properties for a general motion propagator with resets from a mesoscopic perspective. From all the existing papers, in \cite{EuMe16} Eule and Metzger perform a similar study to ours but using Langevin dynamics to describe the movement. Our work differs from that one in the fact that we start from a general motion propagator $P(x,t)$, which allows us to derive an elegant and treatable expression for the first moment and the MSD of the overall process in terms of the motion first moment and MSD respectively (see Eq. \eqref{Eq3}). Moreover, the formalism herein employed eases the inclusion of processes which are not trivial to model in the Langevin picture as L\'evy flights or L\'evy walks.

This paper is organized as follows. In Section \ref{SecMSD} we find an expression for the propagator of the overall process in the Laplace-position space and a general formula for the MSD of the overall process in terms of the motion MSD; the first arrival properties of the system are studied in Section \ref{SecFAT}. In Section \ref{SecCTRW} we apply the general results to three types of movement (sub-diffusive, diffusive and L\'evy) and in Section \ref{SecRWPot} we apply the formalism to study the transport properties and the first arrival of a biased Brownian oscillator. Finally, we conclude the work in Section \ref{SecConclusions}.
\section{General Formulation}
\label{SecGeneral}
In this section we use a renewal formalism to study both the transport properties of a random motion and its first arrival statistics. Concretely, we derive formulas for the global properties of the system in terms of the type of random motion and the reset distribution. We focus in three measures which are of special interest in the study of movement processes: the first moment, the MSD and the mean first arrival time (MFAT).
\subsection{Transport properties}
\label{SecMSD}
Let us consider a general motion propagator $P(x,t)$ starting at $x=0$ and $t=0$ which is randomly interrupted and starts anew at times given by a reset time distribution $\varphi_R(t)$. When one of these reset happens, the motion instantaneously recommences from $x=0$ according to $P(x,t)$ and so on and so forth. Then, the propagator of the overall process, which we call $\rho(x,t)$, is an iteration of multiple repetitions of $P(x,t)$ and the running time of each is determined by $\varphi_R(t)$. 

We start by building a mesoscopic balance equation for $\rho(x,t)$. For simplicity, we assume that the overall process starts at the origin. Then, the following integral equation is fulfilled:
\begin{equation}
\rho(x,t)=\varphi_R^*(t)P(x,t)+\int_0^t \varphi_R(t')\rho(x,t-t') dt',
\label{Eq1}
\end{equation}
where $\varphi_R^*(t)=\int_t^{\infty}\varphi_R(t')dt'$ is the probability of the first reset happening after $t$. The first term in the r.h.s. accounts for the cases where no reset has occurred until $t$ and, therefore, the overall process is described by the motion propagator. The second term accounts for the cases where at least one reset has occurred before $t$ and the first one has been at time $t'<t$, in which case the system is described by the overall propagator with a delay $t'$. Notably, we have introduced $\rho(x,t-t')$ as the propagator of the process starting at $x=0$ at time $t'$ (formally, it should be $\rho(x,t;0,t')$). This can be done independently of the form of $P(x,t)$ as long as the first realization of the process does not affect the following ones. When this is so, the scenario at $t'$ is equivalent to a system starting at $t_0=0$ and having a time $t-t'$ to reach $x$. 

Taking Eq. \eqref{Eq1} to the Laplace space for the time variable, we can isolate the propagator of the overall process to be
\begin{equation}
\hat{\rho}(x,s)=\frac{\mathcal{L}\left[ \varphi_R^*(t)P(x,t)\right]}{1-\hat{\varphi}_R(s)}
\label{Eq2}
\end{equation}
where $\mathcal{L}[f(t)]=\hat{f}(s)=\int_0^\infty e^{-st}f(t)dt$ denotes the Laplace transform. We can now obtain a general equation for the first moment of the overall process multiplying by $x$ at both sides of Eq. \eqref{Eq2} and integrating over $x$. Doing so, one gets
\begin{equation}
\langle \hat{x}(s)\rangle=\frac{\mathcal{L}\left[ \varphi_R^*(t)\langle x(t)\rangle_m \right]}{1-\hat{\varphi}_R(s)},
\label{Eq3_0}
\end{equation}
where $\langle x(t)\rangle_m$ is the time-dependent first moment of the motion process. Nevertheless, usually this process is symmetric and its first moment is zero. In these cases, the second moment or MSD becomes the most relevant magnitude to describe the transport of the system. From Eq. \eqref{Eq2}, instead of multiplying by $x$, if we do so by $x^2$ and integrate over $x$ we get
\begin{equation}
\langle \hat{x}^2(s)\rangle=\frac{\mathcal{L}\left[ \varphi_R^*(t)\langle x^2(t)\rangle_m \right]}{1-\hat{\varphi}_R(s)},
\label{Eq3}
\end{equation}
where $\langle x^2(t)\rangle_m$ is the motion MSD. The importance of this equation lies in the fact that, if we know the motion MSD and the reset time probability density function (PDF) separately, we can introduce them into Eq. \eqref{Eq3} and directly obtain the transport information about the overall process.  

The renewal formulation used herein differs from the method most commonly used in the bibliography to study random walk processes with resets, consisting on introducing a reset term ad-hoc to the Master equation of the process (see \cite{EvMa11} for instance). Contrarily, it resembles the techniques employed in \cite{KuGu15} to study L\'evy flights with exponentially distributed resets or in \cite{PaRe17} to study from a general perspective the first passage problem with resets. In these works, processes described by a known propagator or completion time distribution which are subject to resets are studied using a renewal approach.

\subsubsection{Exponentially distributed reset times}
Let us study the particular case where reset times are exponentially distributed ($\varphi_R(t)=\frac{1}{\tau_m}e^{-t/{\tau_m}}$), keeping the movement as general as before. In this scenario, the real space-time propagator of the overall process in Eq. \eqref{Eq2} can be found by applying the inverse Laplace transform to be
\begin{equation}
\rho_{e}(x,t)= e^{-\frac{t}{\tau_m}}P(x,t)+\frac{1}{\tau_m}\int_0^t e^{-\frac{t'}{\tau_m}}P(x,t') dt'.
\label{Eq6}
\end{equation}
Under the condition that the Laplace transform of $P(x,t)$ exists at $s=\frac{1}{\tau_m}$, an equilibrium is reached and the distribution there can be generally written as
\begin{equation}
\rho_{e}(x)=\frac{\hat{P}(x,\frac{1}{\tau_m})}{\tau_m}.
\label{Eq7}
\end{equation}
The required condition for the equilibrium distribution to exist includes a wide range of processes from the most studied in the bibliography: Brownian motion, L\'evy flights, etc. Similarly, an expression for the equilibrium first moment of the overall process in terms of the motion first moment can be derived from Eq. \eqref{Eq3_0} reading
\begin{equation}
\langle x\rangle_{e}(\infty)=\frac{\langle \hat{x}(\frac{1}{\tau_m})\rangle_m}{\tau_m},
\label{Eq10_0}
\end{equation}
and for the MSD we have:
\begin{equation}
\langle x^2\rangle_{e}(\infty)=\frac{\langle \hat{x^2}(\frac{1}{\tau_m})\rangle_m}{\tau_m}.
\label{Eq10}
\end{equation}

Eq. \eqref{Eq10} introduces an extra condition on the type of motion for it to define a finite area around the origin: the Laplace transform of its MSD must be finite at $s=\frac{1}{\tau_m}$. For instance, despite L\'evy flights reach an equilibrium state when they are subject to constant rate resets, since its MSD diverges so does the MSD of the overall process. 

Multiple processes can be found in the bibliography with a MSD which is Laplace transformable and, therefore, reach an equilibrium MSD when exponential resets are applied to them. Some of these processes are L\'evy walks, ballistic or even turbulent motion \cite{KlBl87}. Notably, Eq. \eqref{Eq7} is also applicable to movement in more than one dimension when it is rotational invariant. In this case, the movement can be described by a one-dimensional propagator $P(r,t)$ where $r$ is the radial distance from the origin. Therefore, any process without a preferred direction as correlated random-walks and L\'evy walks in the plane \cite{BaDa05} or self-avoiding random walks for arbitrary spatial dimension \cite{DuHa13} are significant processes which form a finite size area when they are subject to exponential resets.

\subsection{First arrival}
\label{SecFAT}
The second remarkable result from \cite{EvMa11} is the existence of a finite MFPT when diffusive motion is subject to constant rate resets. Since then, several works have been published focused on the first completion time with resets \cite{RoRe15,Re16,PaRe17} but none of them have put the focus on the generality of these results with respect to the properties of the random motion. During the writing of this paper we have realized that a deep analysis of the first passage for search processes has been recently done in \cite{ChSo18}. Nevertheless, besides our general qualitative analysis is similar to the one performed there, we study in detail cases of particular interest in a movement ecology context as sub-diffusive motion, L\'evy flights or random walks in potential landscapes. Moreover, we perform numerical simulations of the process to check our analytical results.

In this work we use the MFAT as a measure of the time taken by the process to arrive to a given position, instead of crossing it as is considered in the MFPT. This is motivated by the fact that for L\'evy flights the MFPT has an ambiguous interpretation due to the possibility of extremely long jumps in infinitely small time steps. Contrarily, the MFAT can be clearly interpreted and its properties have been deeply studied in \cite{ChMe03}. 

Before focusing on practical cases, let us start with the general renewal formulation. We build a renewal equation for the survival probability of the overall process $\sigma_x(t)$ in terms of the survival probability of the motion $Q_x(t)$ and the reset time PDF $\varphi_R(t)$, similar to the equation for the propagator in the previous section:
\begin{equation}
\sigma_x(t)=\varphi_R^*(t)Q_x(t)+\int_0^t \varphi_R(t')Q_x(t')\sigma_x(t-t')dt'.
\label{Eq12}
\end{equation}
Here, the first term on the r.h.s. corresponds to the probability of not having reached $x$, nor a reset has occurred in the period $t\in (0,t]$. The second term is the probability of not having reached $x$ when at least one reset has happened at time $t$. In the latter, we account for the probability $Q_x(t')$ of not having reached $x$ in the first trip, which ends at a random time $t'$, and the probability of not reaching $x$ at any other time after the first reset $\sigma_x(t-t')$; and these two conditions are averaged over all possible first reset times $t'$. Applying the Laplace transform and isolating the overall survival probability we obtain:
\begin{equation}
\hat{\sigma}_x(s)=\frac{\mathcal{L}[\varphi_R^*(t) Q_x(t)]}{1-\mathcal{L}[\varphi_R(t) Q_x(t)]}.
\label{Eq13}
\end{equation}
This equation, which has been recently derived by similar means in \citep{ChSo18}, is the cornerstone from which the existence of the MFAT is studied. If in the asymptotic limit the survival probability behaves as 
\begin{equation}
\sigma_x(t)\sim t^{-\beta},
\end{equation}
for $\beta>1$ the MFAT is finite, while for $\beta \leq 1$ it diverges. Since we have the expression of the survival probability in the Laplace space, it is convenient to rewrite these conditions for $\hat{\sigma}_x(s)$ instead. Let us consider the following situations:

i) When $\beta >1$, the Laplace transform of the survival probability tends to a constant value for small $s$. The MFAT is finite and can be found as 
\begin{equation}
T_{F}=\int_0^\infty t q_x(t)dt= \lim_{s\rightarrow 0} \hat{\sigma}_x(s)
\label{Eq14}
\end{equation}
where $q_x(t)=-\frac{\partial \sigma_x(t)}{\partial t}$ is the first arrival time distribution of the overall process. Concretely, the MFAT can be found in terms of the distributions defined above as
\begin{equation}
T_F=\frac{\int_0^{\infty}\varphi_R^*(t)Q_x(t)dt}{1-\int_0^{\infty}\varphi_R(t)Q_x(t)dt}.
\label{Eq15}
\end{equation}

ii) When $\beta =1$, the Laplace transform of the survival probability tends to infinity for small $s$. Therefore, in this case, the MFAT is infinite since
\[
\lim_{s\rightarrow 0} \hat{\sigma}_x(s)=\infty
\]
so
\[
T_F =\infty
\]

iii) When $\beta < 1$, the Laplace transform of the survival probability diverges as $\hat{\sigma}_x(s)\sim s^{\beta-1}$ for small $s$. The MFAT is infinite and the survival probability decays as $\sigma_x(t)\sim t^{-\beta}$ with time.

Notably, when the reset times are exponentially distributed ($\varphi_R(t)=\frac{1}{\tau_m}e^{-\frac{t}{\tau_m}}$), the MFAT of the overall process is always finite for motion survival probabilities which are Laplace-transformable. Concretely, in this particular case Eq. \eqref{Eq15} reduces to
\begin{equation}
T_{F}=\frac{\tau_m\hat{Q}_x(\frac{1}{\tau_m})}{\tau_m-\hat{Q}_x(\frac{1}{\tau_m})}.
\label{Eq16}
\end{equation}
\section{Free motion}
\label{SecCTRW}
In order to get a deeper intuition about the results in the previous section, let us take generic expressions for both the reset time distribution and the motion propagator. In the first place we study well-known processes which do not have environmental constrains (potential landscapes, barriers, etc.). 
\subsection{Transport properties}
\begin{figure}
    \centering
        \includegraphics[scale=0.55]{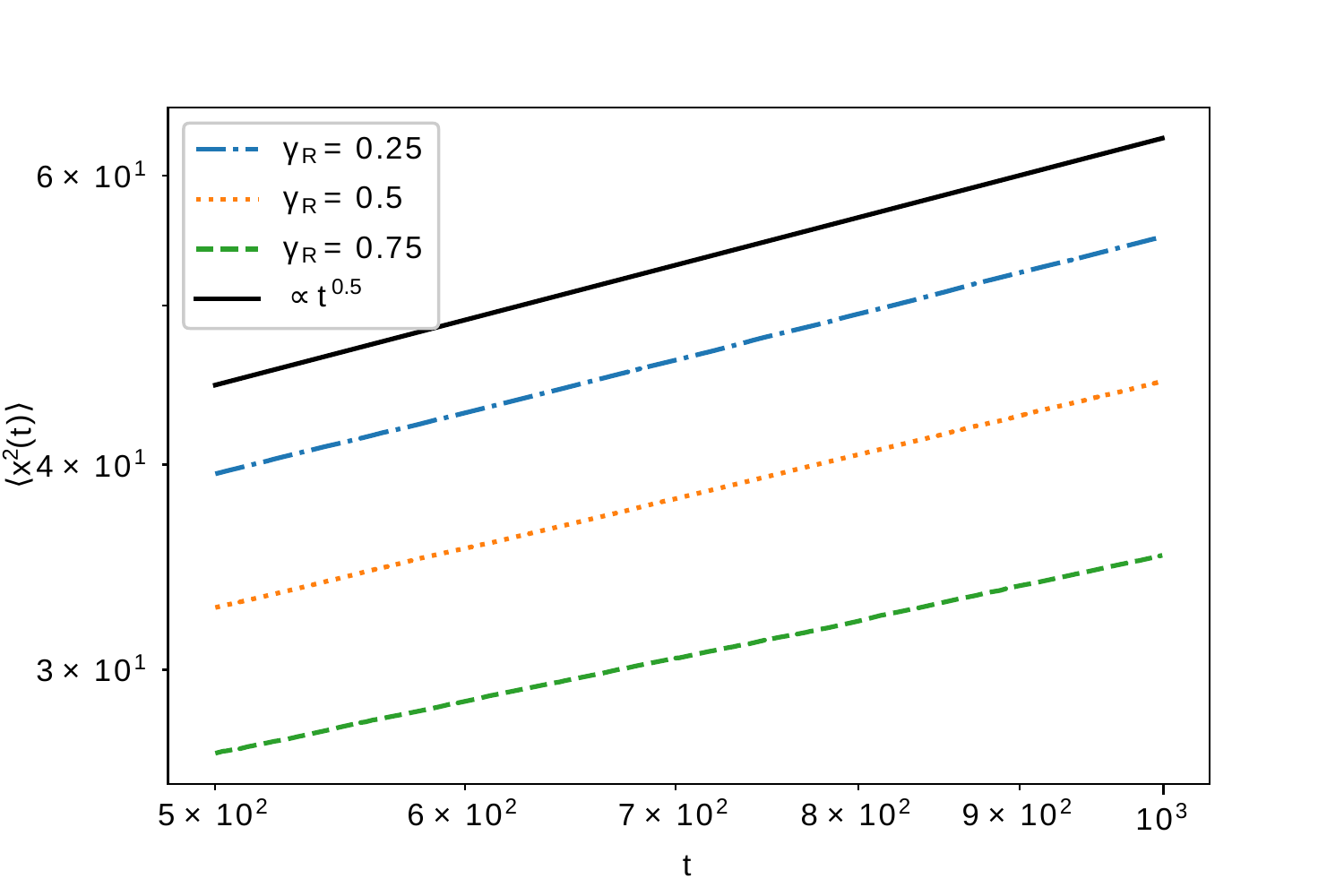}
    \caption{The asymptotic behaviour of the MSD of the overall process for sub-diffusive motion with $p=0.5$ is shown in a log-log plot. Three different exponents $\gamma_R<1$ for the reset time distribution are considered and all of them are seen to scale as the solid black guide line of slope $0.5$. Therefore, the reset exponent $\gamma_R$ only affects multiplicatively to the transport regime.}
    \label{Fig1}
\end{figure}

Let us start by studying the transport properties of the overall process for a symmetric motion, i.e.

\begin{equation}
\langle x(t)\rangle_m =0,
\label{mean}
\end{equation}
with a MSD scaling as
\begin{equation}
\langle x^2(t)\rangle_m \sim t^p,
\label{mdp}
\end{equation}
with $0<p\leq 2$. This choice includes sub-diffusive motion for $p<1$, diffusive motion for $p=1$ and super-diffusive motion with for $p>1$. Also, we take the reset time distributions to be
\begin{equation}
\varphi_R(t)=\frac{t^{\gamma_R-1}}{\tau_m^{\gamma_R}}E_{\gamma_R,\gamma_R}\left[-\left( \frac{t}{\tau_m}\right)^{\gamma_R} \right],
\label{Eq4}
\end{equation}
with $0<\gamma_R \leq 1$, where $$E_{\alpha,\beta}(z)=\sum_{n=0}^{\infty}\frac{(-z)^{n}}{\Gamma(\alpha n +\beta)}$$ is the generalized Mittag-Leffler function with constant parameters $\alpha$ and $\beta$. This allows us to recover the exponential distribution for $\gamma_R=1$ and we can also study power law behaviors of the type $\varphi_R(t)\sim t^{-1-\gamma_R}$ for $\gamma_R <1$. For this distribution, the survival probability $\varphi_R^*(t)=\int_t^\infty\varphi(t')dt'$ reads:
\begin{equation}
\varphi_R^*(t)=E_{\gamma_R,1}\left[-\left( \frac{t}{\tau_m}\right)^{\gamma_R} \right].
\label{Eq4cum}
\end{equation}
For a wide study about the properties of the Mittag-Leffler function we refer the reader to \cite{MLFunctions}. In this case, since the first moment is zero, the MSD becomes the most significant moment of the process. From Eq. \eqref{Eq10} one can see that MSD of the overall process has two possible behaviors for large $t$ (see Appendix A1 for details):
\begin{equation}
\langle x^2(t)\rangle \sim \left\{
\begin{matrix} t^p,\ \text{for}\ \gamma_R<1\\
{}\\
t^0,\ \text{for}\ \gamma_R=1\\
\end{matrix} \right. .
\label{Eq5}
\end{equation}
Therefore, for power law reset time PDFs with any exponent $\gamma_R <1$, the MSD of the overall process scales as the motion MSD, so that a long-tailed reset PDF does not modify the transport regime. To illustrate this, in Fig. \ref{Fig1} we show simulations of the asymptotic behavior of the overall MSD for sub-diffusive motion with long-tailed resets. There we see that, as can be seen from Eq. \eqref{Eq5}, long-tailed resets only affect the transport multiplicatively but not modify the regime.

When the motion is a L\'evy flight, the MSD diverges for all $t$, i.e. $$\langle x^2(t)\rangle_m=\infty, \ t>0.$$ Hence, from Eq. \eqref{Eq3}, the MSD of the overall process also diverges for any reset time PDF except for the pointless case $\varphi_R(t)=\delta(t)$.

\begin{figure}
    \centering
        \includegraphics[scale=0.6]{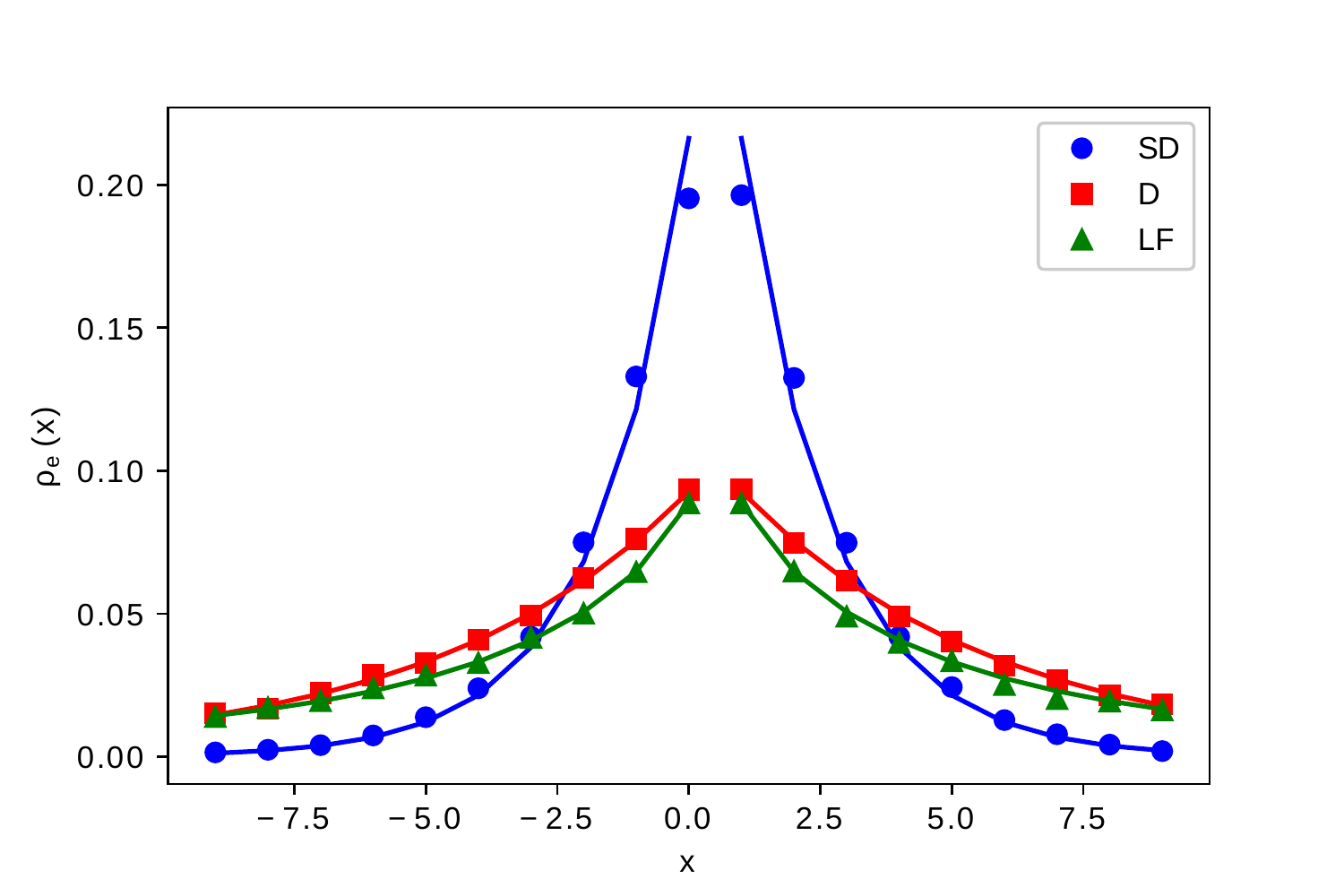}
    \caption{Equilibrium distribution of the overall process with sub-diffusive (SD) with $\gamma =0.5$, diffusive (D) and L\'evy flight (LF) with $\alpha =1.5$ motion propagator, all with $D=0.1$, and exponential reset times with $\tau_m=10$. Each stochastic simulation is compared to the corresponding analytical expressions \eqref{Eq8} and \eqref{Eq9} (solid lines).}
     \label{FigStatDists}
\end{figure}

Regarding exponential reset time distributions case ($\gamma_R=1$), an equilibrium state is reached and we can in principle compute an equilibrium distribution. We start by considering a sub-diffusive propagator (see Eq. (42) in \cite{MeKl00}) which, in the Fourier-Laplace space, reads:
\begin{equation}
\hat{\tilde{P}}(k,s)=\frac{1}{s+D s^{1-\gamma}k^2},
\label{EqSubForProp}
\end{equation}
with $D$ the (sub-)diffusion constant. This propagator describes sub-diffusive movement for $\gamma <1$ and diffusive movement for $\gamma =1$. Then, the equilibrium distribution given by Eq. \eqref{Eq7} becomes a symmetric exponential distribution 
\begin{equation}
\rho_{e}(x)=\frac{1}{\sqrt{4 D \tau_m^\gamma}}e^{-\frac{|x|}{\sqrt{D \tau_m^\gamma}}},
\label{Eq8}
\end{equation}
where for $\gamma=1$ we recover the equilibrium distribution found in \cite{EvMa11}. If instead of a sub-diffusive propagator, we consider a super-diffusive motion and, in particular, the propagator for a L\'evy flight in the Fourier-Laplace space
\begin{equation}
\hat{\tilde{P}}(k,s)=\frac{1}{s+D |k|^\alpha}
\end{equation}
with $\alpha<2$ and $D$ a constant, the equilibrium distribution of the overall process becomes:
\begin{equation}
\rho_{e}(x)=2\int_0^\infty \frac{\cos(kx)}{1+\tau_m D k^\alpha}dk.
\label{Eq9}
\end{equation}
In Fig. \ref{FigStatDists} we compare both analytical results in Eq. \eqref{Eq8} and Eq. \eqref{Eq9} with numerical Monte-Carlo simulations of the process. The agreement is seen to be excellent.
\subsection{First Arrival}
\begin{figure}
\centering
        \includegraphics[scale=0.6]{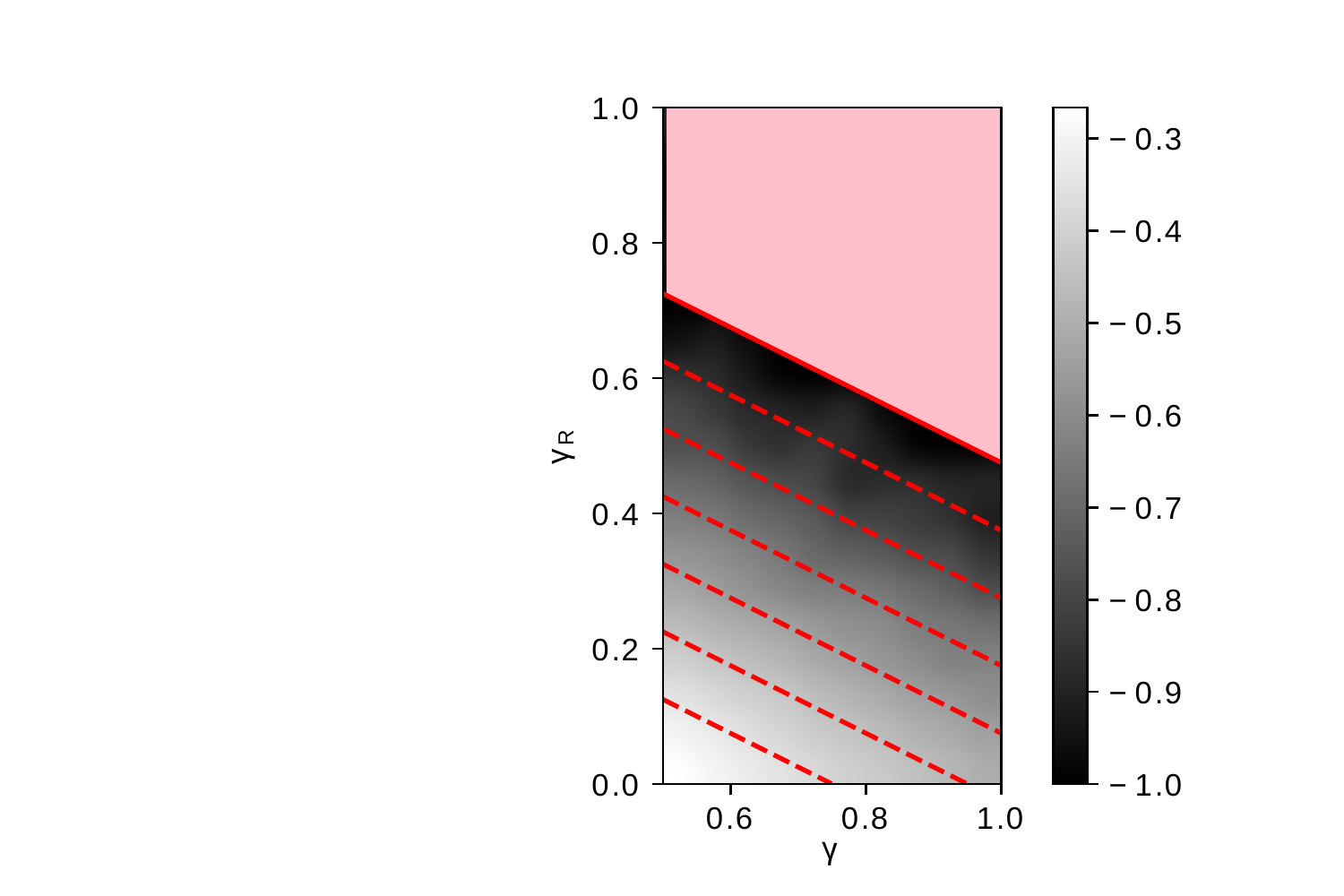}
                ~
        \includegraphics[scale=0.6]{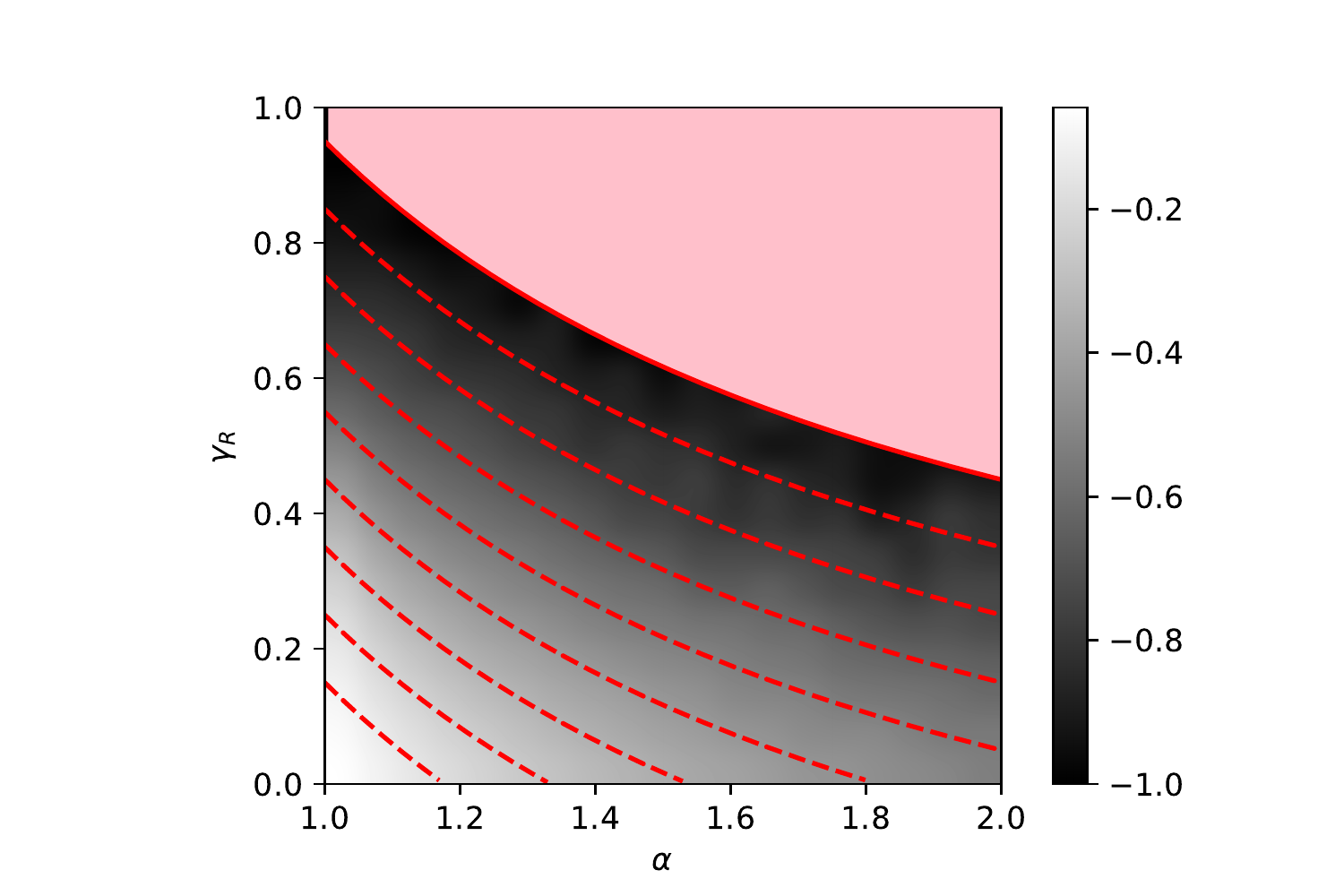}
    \caption{The tail exponent $\beta$ of the survival probability $\sigma_x(t)\sim t^{\beta}$ of the overall process at $x=0.5$ is shown for sub-diffusive motion with exponent $\gamma$ in A and L\'evy flight motion with exponent $\alpha$ in B, both subject to resets at times given by a Mittag-Leffler distribution with tail parameter $\gamma_R$ and $\tau_m =10$. The values of $\beta$ have been computed for $\gamma_R\in \{0.05,0.1...0.9,0.95\}$, and $\gamma\in \{0.5,0.55...0.9,0.95\}$ in panel A and $\alpha\in \{1.05,1.1...1.9,1.95\}$ in panel B. Gaussian interpolation has been applied in order to smooth the simulated results. In each plot, the solid red curve ($\gamma_R+\frac{\gamma}{2}=1$ in A and $\gamma_R-\frac{1}{\alpha}=0$ in B) corresponds to the limit between finite (flat pink region) and infinite (gradient) MFAT. The dashed curves are the analytical level curves for $\beta = 0.9,0.8...$ from top to bottom. The black regions observed just below the limiting curves are due to the discretization of the parameter space.}
    \label{FigMFAT}
\end{figure}

Let us now study the MFAT for a general motion survival probability decaying as
\begin{equation}
Q_x(t)\sim t^{-q}, \quad q>0
\label{surv}
\end{equation}
for long $t$ and the same reset time distribution defined in Eq. \eqref{Eq4}. Under these assumptions, the asymptotic behavior of the overall survival probability is (see Appendix A2 for details)
\begin{equation}
\sigma_x(t) \sim t^{-\gamma_R-q}, \quad \text{if}\ \gamma_R+q\leq 1,
\label{Eq14}
\end{equation}
as has been recently found in \cite{ChSo18} by similar means. This implies that, in this case, $T_{F}=\infty$. However, when $\gamma_R+q>1$ the MFAT is finite and can be expressed as
\begin{equation}
T_{F}(x)=\frac{\int_0^\infty E_{\gamma_R,1}\left[-\left( \frac{t}{\tau_m}\right)^{\gamma_R} \right] Q_x(t)dt}{1-\int_0^\infty \frac{t^{\gamma_R-1}}{\tau_m^{\gamma_R}}E_{\gamma_R,\gamma_R}\left[-\left( \frac{t}{\tau_m}\right)^{\gamma_R} \right] Q_x(t) dt}.
\label{Eq3_11}
\end{equation}
The two regions where the MFAT is finite and infinite for a sub-diffusive (Fig. \ref{FigMFAT} A) and a L\'evy flight motion process (Fig. \ref{FigMFAT} B) are shown in Fig. \ref{FigMFAT}. Let us study these two cases separately. As shown in \cite{RaDi00}, for a sub-diffusive motion, the survival probability in the long time limit decays as
\begin{equation}
Q_x(t)\sim t^{-\frac{\gamma}{2}}
\label{EqSDSurv}
\end{equation}
with $0<\gamma <1$. For $\gamma=1$ we recover the survival probability of a diffusion process. Here we can identify $q=\frac{\gamma}{2}$ and from Eq. \eqref{Eq14} the survival probability of the overall process decays as
\begin{equation}
\sigma_x(t) \sim t^{-\gamma_R-\frac{\gamma}{2}},
\label{EqSDSurvOAll}
\end{equation}
when $\gamma_R+\frac{\gamma}{2}\leq 1$ and the MFAT is infinite in this region of exponents. Contrarily, the MFAT is finite when $\gamma_R+\frac{\gamma}{2}>1$. This result has been compared with stochastic simulations of the process (Fig. \ref{FigMFAT}A), where the limiting curve $\gamma_R=1-\frac{\gamma}{2}$ and the tail exponent for the overall survival probability are in clear agreement with the analytical results. 

We have also studied the survival probability when the underlying motion is governed by L\'evy flights propagator. In this case, the survival probability decays as
\begin{equation}
Q_x(t)\sim t^{\frac{1}{\alpha}-1},
\label{EqLFSurv}
\end{equation}
with $1<\alpha<2$, as shown in \cite{ChMe03}. Here we can also recover the diffusive behaviour for $\alpha=2$. Identifying $q=1-\frac{1}{\alpha}$, the overall survival probability reads
\begin{equation}
\sigma_x(t) \sim t^{\frac{1}{\alpha}-\gamma_R-1}
\label{EqLFSurvOAll}
\end{equation}
in the asymptotic limit and for $\gamma_R-\frac{1}{\alpha}\leq 0$. In this case the MFAT is infinite while for $\gamma_R-\frac{1}{\alpha}>0$ it is finite. In Figure \ref{FigMFAT}B we present the results to see that these two regions are also found in a stochastic simulation of the overall process.

\begin{figure}
    \centering
        \includegraphics[scale=0.6]{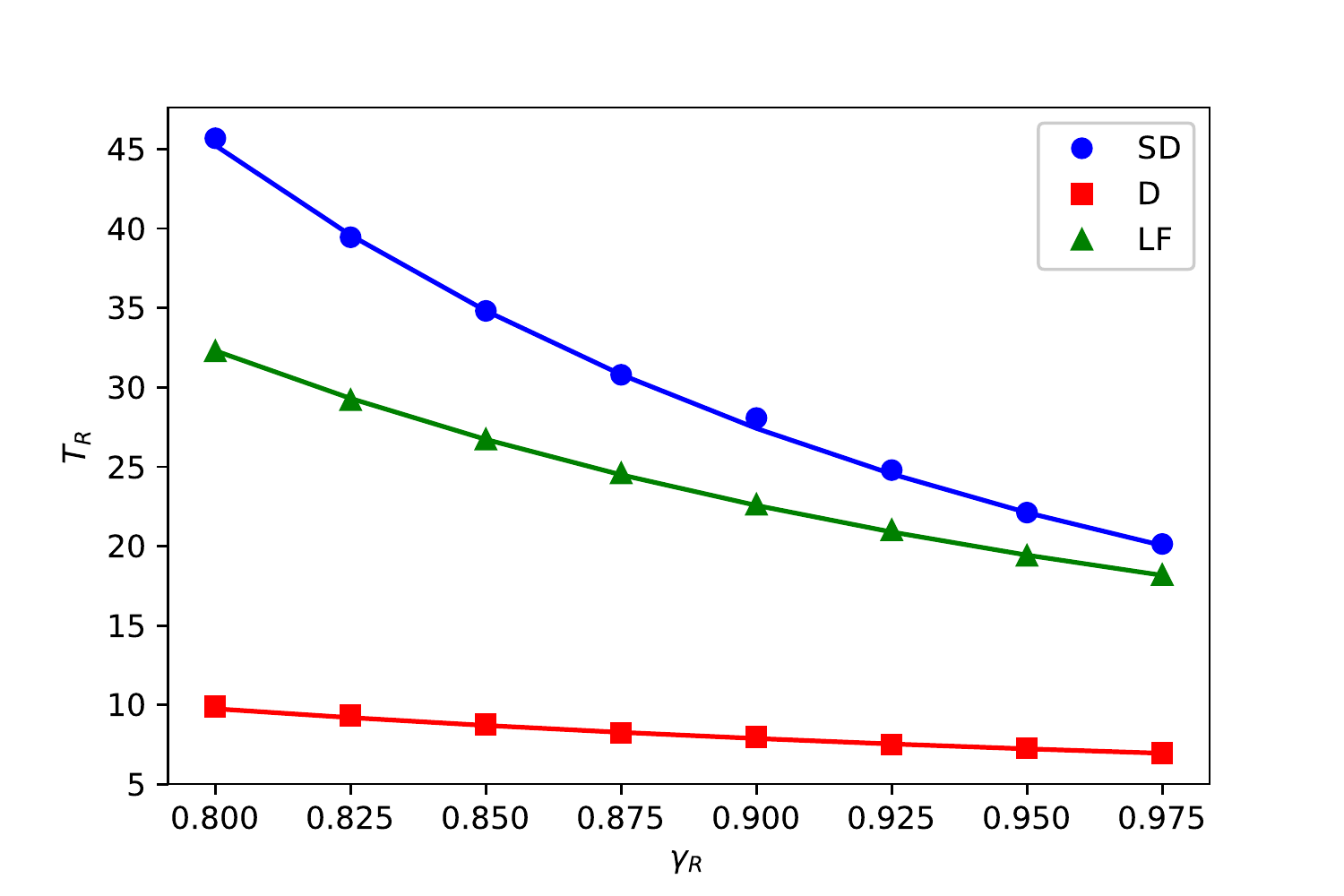}
    \caption{The simulated MFAT for three representative cases of sub-diffusion (SD) with $\gamma=0.5$, diffusion (D) with $\gamma=1$ and a L\'evy Flight with $\alpha=1.5$, all with $D=0.1$, are compared to the analytical results obtained from Eq. \eqref{Eq15} (solid curves) for different reset distribution tail exponents $\gamma_R$ and $\tau_m=10$. Concretely, the MFAT is computed at a distance $x=0.5$ from the origin.}
     \label{FigMFAT2}
\end{figure}

Unlike the existence of an equilibrium MSD, the finiteness of the MFAT is not drastically broken when the reset time distribution changes from short to long-tailed. A remarkable property that we can see in Fig. \ref{FigMFAT} is that both the reset time distribution and the motion first arrival time distribution can have an infinite mean value and still the mean value of the overall process is finite. This property has been explicitly tested by computing the simulated MFAT for parameters in the white region in Fig. \ref{FigMFAT} for both the sub-diffusive and the L\'evy flight case, and also for the diffusive limiting case. The simulated MFAT is compared to the one obtained from numerical integration of Eq. \eqref{Eq15} and the results are shown to be in agreement (Fig. \ref{FigMFAT2}). 
\section{Brownian motion in a biased harmonic potential}
\label{SecRWPot}

In this section we study the transport and the first arrival statistics of a Brownian particle starting at $x=0$ with a white, Gaussian noise with diffusion constant $D$. It moves inside a biased harmonic potential $V(x)=\frac{1}{2}k(x-x_0)^2$ and it has a drift $\gamma$. Unlike the cases studied in the previous section, here the movement has an intrinsic bias towards the point $x_0$. This, in an ecological context, can be seen as the knowledge the animal have about the optimal patches to find food.

\begin{figure}
\includegraphics[scale=0.6]{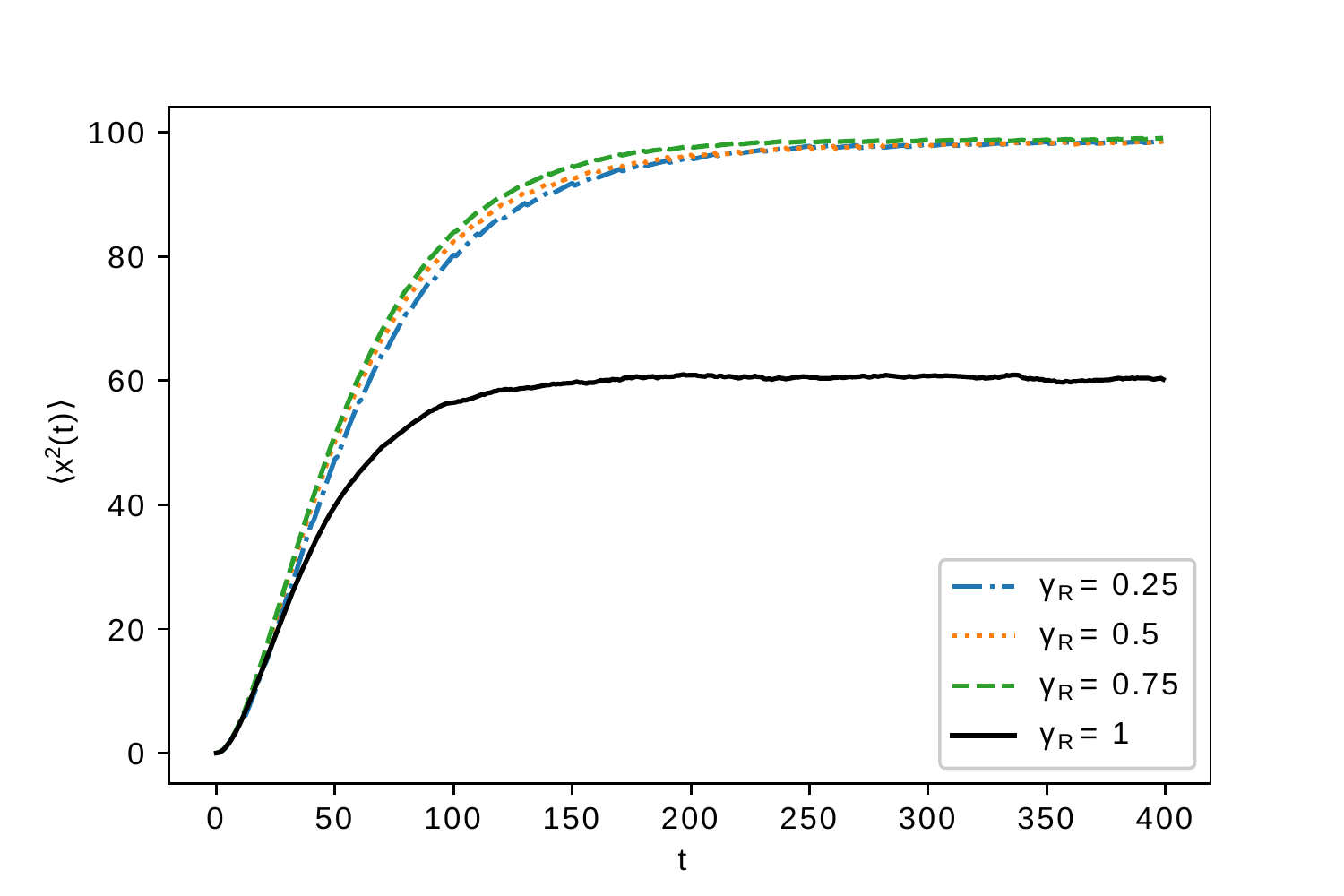}
\includegraphics[scale=0.6]{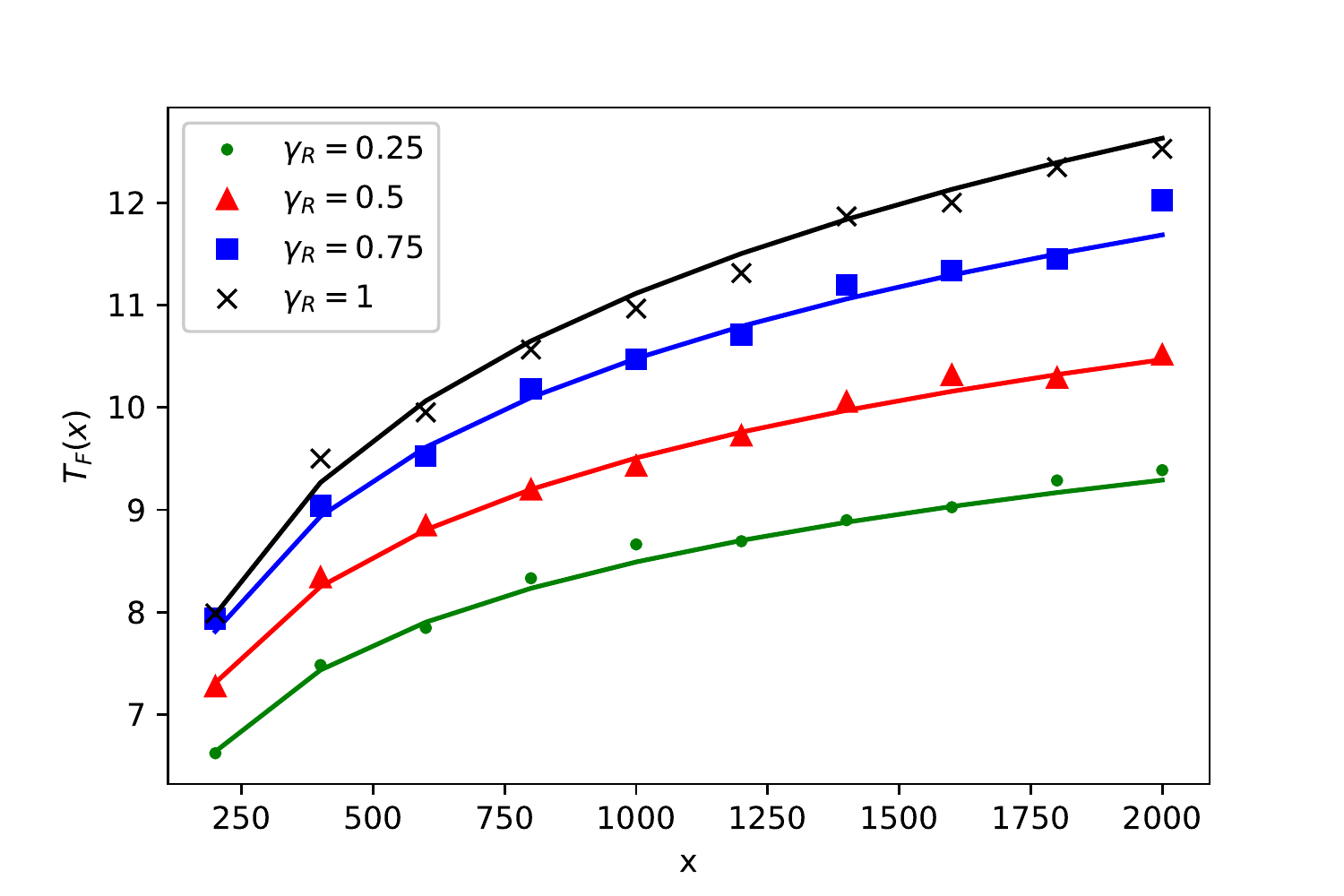}
\caption{In panel A, we show the MSD as a function of time for a simulation of a Brownian motion ($D=0.25$) with an harmonic force $F=1$ with equilibrium point at $x_0=10$ subject to resets with $\tau_m=10$ and different $\gamma_R$ parameters. For all three $\gamma_R<1$ the MSD tends to the same value, which is larger than the equilibrium MSD for $\gamma_R=1$. Below, in panel B, the simulated MFAT for $F=1$, $D=1$ and $\tau_m=10$ is compared to the analytical result for Mittag-Leffler reset distributions with different $\gamma_R$ (solid curves).}
\label{FigMSDpot}
\end{figure}

When this system is constrained by constant rate resets (i.e. exponentially distributed reset times), an equilibrium distribution is attained as shown in \cite{Pa15} by introducing resets to the Fokker-Plank equation of the system. Instead, with the general formalism derived in Section \ref{SecGeneral}, we can first study the system without resets and introduce the results to the formulas derived above. Then, we start from the Langevin equation for the Brownian particle in an harmonic potential (\cite{HuCh10}):
\begin{equation}
\frac{dx}{dt}= -\frac{D}{\gamma}\frac{\partial V(x)}{\partial x}+\sqrt{2D}\eta(t),
\label{Eq4_1}
\end{equation}
where the over-dumped limit has been implicitly taken and $\eta(t)$ is a Gaussian noise so that $\langle \eta(t)\rangle=0$ and $\langle \eta(t)\eta(t')\rangle =\delta(t-t')$ (i.e. a white noise). For a biased harmonic potential it becomes
\begin{equation}
\frac{dx}{dt}=-D F (x-x_0)+\sqrt{2D}\eta(t)
\label{Eq4_2}
\end{equation}
where $F=k/\gamma$ has been defined. From this equation, the first moment and the MSD of the particle can be derived (see \cite{Ga09} for specific methods and tools) to be
\begin{equation}
\langle x(t)\rangle_m=x_0\left( 1-e^{-DFt}\right)
\label{Eq4_3_0}
\end{equation}
and
\begin{equation}
\begin{split}
\langle x^2(t)\rangle_m=&\left( \frac{1}{F}+x_0^2\right) \left( 1-e^{-2DFt}\right)\\
&-2x_0^2e^{-DFt}\left( 1-e^{-DFt}\right)
\end{split}
\label{Eq4_3}
\end{equation}
respectively. Introducing these expressions to the main equations for the moments of the process (Eq. \eqref{Eq3_0} and Eq. \eqref{Eq3}) we can obtain the Laplace space dynamics of the mean 
\begin{equation}
\langle\hat{x}(s)\rangle=x_0\left(\frac{1}{s}-\frac{\hat{\varphi}_R^*(s+DF)}{1-\hat{\varphi}_R(s)}\right)
\end{equation}
and the MSD
\begin{equation}
\begin{split}
\langle\hat{x}^2(s)\rangle=&\left( \frac{1}{F}+x_0^2\right)\left(\frac{1}{s}-\frac{\hat{\varphi}_R^*(s+2DF)}{1-\hat{\varphi}_R(s)}\right)\\
&-2x_0^2\frac{\hat{\varphi}_R^*(s+DF)-\hat{\varphi}_R^*(s+2DF)}{1-\hat{\varphi}_R(s)}.
\end{split}
\end{equation}
in terms of the reset time PDF. For small $s$, the terms with $1-\hat{\varphi}_R(s)$ in the denominator can be neglected w.r.t. the term $\frac{1}{s}$ when the distribution $\varphi_R(t)$ is long tailed. This is because in the $s\rightarrow 0$ limit, the numerator of these terms remains finite while the denominator $1-\hat{\varphi}_R(s)\sim s^{\gamma_R}$, with $\gamma_R<1$. Therefore, for long tailed reset time distributions, the first is the dominant term. On the other hand, for exponentially distributed resets we have that
$$
\frac{\hat{\varphi}_R^*(s+2DF)}{1-\hat{\varphi}_R(s)}=\frac{1}{s}\frac{1}{1+2DF\tau_m}+\mathcal{O}(s^0)
$$
and, equivalently, 
$$
\frac{\hat{\varphi}_R^*(s+DF)}{1-\hat{\varphi}_R(s)}=\frac{1}{s}\frac{1}{1+DF\tau_m}+\mathcal{O}(s^0).
$$
Therefore, the equilibrium first moment and MSD of the overall process can be seen to be
\begin{equation}
\langle x(\infty)\rangle_e=x_0\left( 1-\kappa_{\gamma_R}\frac{1}{1+DF\tau_m}\right)
\label{Eq4_7}
\end{equation}
and
\begin{equation}
\begin{split}
\langle x^2(\infty)\rangle_e=&\left(\frac{1}{F}+x_0^2\right)\left( 1-\kappa_{\gamma_R}\frac{1}{1+2DF\tau_m}\right)\\
&-\kappa_{\gamma_R}2x_0^2 \frac{DF\tau_m}{(1+DF\tau_m)(1+2DF\tau_m)}
\end{split}
\label{Eq4_8}
\end{equation}
respectively, with $\kappa_{\gamma_R}=1\ \text{for}\ \gamma_R=1$ and $\kappa_{\gamma_R}=0\ \text{for}\ \gamma_R<1$. Then, when the reset distribution is long-tailed, both the equilibrium mean and MSD are equal to the ones for the process without resets. However, when resets are exponentially distributed, the values of the equilibrium mean and MSD are diminished by the factors preceeded by $\kappa_{\gamma_R}$ in the equations right above. This difference has been tested for the MSD by means of a stochastic simulation of the Langevin equation (Fig. \ref{FigMSDpot}). As happens for the transport properties of the free processes studied in Section \ref{SecCTRW}, for this type of movement we also find that long-tailed reset distributions do not affect the significant features of the MSD (and also the mean in this case), while reset times which are distributed exponentially do affect activelly the long time behavior of the overall process.

Let us now study its MFAT for this system. The first arrival distribution $q_{x_0}(t)$ at the minimum of the potential for this motion process has been found to be (\cite{HuCh10})
\begin{equation}
q_{x_0}(t)=\frac{2De^{-DFt}|x_0|}{\sqrt{2\pi \sigma_t^3}}\exp\left\{-\frac{(x_0 e^{-DFt})^2}{2\sigma_t}\right\},
\label{Eq4_6}
\end{equation}
from which the survival probability can be found as
\begin{equation}
Q_{x_0}(t)=\int_t^\infty q_{x_0}(t')dt',
\label{Eq4_6p}
\end{equation}
with $\sigma_t=(1-e^{-DFt})/F$. In the asymptotic limit, the first arrival distribution decays as $q_{x_0}(t)\sim e^{-DFt}$ and so does the survival probability $Q_{x_0}(t)\sim e^{-DFt}$ since the decay is exponential. A direct consequence of this is that the global survival probability also has a short tail. This can be seen by looking at Eq. \eqref{Eq13}: when the asymptotic limit of $Q_{x_0}(t)$ is exponential, the expression of the global survival probability in the Laplace space tends to a finite value for small $s$; which is in fact the first arrival time of the global process (see Appendix A3 for further details). 

In Fig. \ref{FigMSDpot} we compare the analytical result predicted by Eq. \eqref{Eq15}, taking the survival probability in Eq. \eqref{Eq4_6} instead of the ones studied in Section \ref{SecCTRW}, with Monte-Carlo simulations of the Langevin equation in Eq. \eqref{Eq4_2}. They are seen to be in perfect agreement. Here, unlike for the free motion processes, resets always penalize the arrival to the target. 

\section{Conclusions}
\label{SecConclusions}
In this work we have derived an expression for the first moment, the MSD and the MFAT of stochastic motion with resets from a unified, renewal formulation. Concretely, we find them in terms of a general resetting mechanism and the type of stochastic motion. This opens the analysis of resets acting on a vast range of stochastic motion processes without the need of building a particular model for each case.

The existence of an equilibrium MSD and a finite MFAT has been tested for a wide class of stochastic processes subject to random resets. The first turns to be extremely sensitive with respect to the reset time distribution. On one hand, when the reset time distribution is long-tailed, the transport regime of the overall process is qualitatively equivalent to the regime of the motion (see Eq. \eqref{Eq5} for free motion and Eqs. \eqref{Eq4_7} and \eqref{Eq4_8} for the biased harmonic Brownian oscillator). On the other hand, for exponential distributions of reset times, qualitative changes are observed regarding the transport of the overall process. Concretely, we have found that for a free motion process with MSD scaling as $\langle x^2(t)\rangle_m \sim t^p$, an equilibrium state with finite MSD is reached. For the Brownian oscillator, both the equilibrium mean and MSD are modified by the resetting mechanism. Therefore, while exponentially distributed resets actively affect the long time behavior of both processes, when long-tailed reset time distributions with infinite mean are chosen, the asymptotics of the motion process are not modified.

Regarding the first arrival time, we have seen that the difference between long-tailed and exponentially distributed reset times is not as marked as for the transport properties. In fact, the transition between them is seen to be soft (compare Fig. \ref{FigMSDpot}A and Fig. \ref{FigMSDpot}B for instance). Interestingly, for the free motion process case, we find that a motion process with an infinite MFAT, when it is restarted at times given by a long-tailed PDF (i.e. with infinite mean), may have a finite MFAT (see Fig. \ref{FigMFAT2})

\section*{Acknowledgements}
This research has been supported by the Spanish government through Grant No. CGL2016-78156-C2-2-R.
\section*{APPENDIX A. ASYMPTOTIC ANALYSIS}
In this appendix we derive the results in Eq. \eqref{Eq5} and Eq. \eqref{Eq14} about the asymptotic behavior of the MSD (A1) and the survival probability (A2) respectively for a motion process with resets. Concretely, we compute them for motion MSD as in Eq. \eqref{mdp} and the survival probability as in Eq. \eqref{surv}. Also, in A3 we compute the MFAT for an exponentially decaying motion survival probability.
\subsection*{A1. Mean square displacement of a free process with resets}
We start by rewriting the general expression for the MSD in Eq. \eqref{Eq3} as
\begin{equation}
\langle \hat{x}(s)\rangle =\frac{T_1(s)}{T_2(s)}
\end{equation}
with $T_1(s)=\mathcal{L}\left[ \varphi_R^*(t)\langle x^2(t)\rangle_P \right]$ and $T_2(s)=1-\hat{\varphi}_R(s)$. In order to study the long $t$ limit of the MSD, in the Laplace space we must study the small $s$ limit. Let us start by $T_2(s)$. In the Laplace space, the Mittag-Leffler distribution can be seen (\cite{MLFunctions}) to be
\begin{equation}
\hat{\varphi}_R(s)=\frac{1}{1+(\tau_m s)^{\gamma_R}}
\end{equation}
from which
\begin{equation}
T_2(s)=1-\frac{1}{1+(\tau_m s)^{\gamma_R}}=\frac{(\tau_m s)^{\gamma_R}}{1+(\tau_m s)^{\gamma_R}}\sim s^{\gamma_R}
\end{equation}
in the small $s$ limit. Let us proceed now with $T_1(s)$. In the long $t$ limit, the Mittag-Leffler survival probability in Eq. \eqref{Eq4cum} can be seen (\cite{MLFunctions}) to behave as
\begin{equation}
\varphi_R^*(t)\sim \left\{
\begin{matrix} t^{-\gamma_R},\ \text{for}\ \gamma_R<1\\
{}\\
e^{-t/\tau_m},\ \text{for}\ \gamma_R=1\\
\end{matrix} \right. .
\label{MLsurvDecay}
\end{equation}
Then, with Eq. \eqref{mdp} it follows that
\begin{equation}
T_1(s)\sim \left\{
\begin{matrix} \mathcal{L}\left[t^{p-\gamma_R} \right]\sim s^{\gamma_R-1-p},\ \text{for}\ \gamma_R<1\\
{}\\
\mathcal{L}\left[t^{p}e^{-t/\tau_m} \right]\sim s^{0},\ \text{for}\ \gamma_R=1\\
\end{matrix} \right. .
\end{equation}
Putting the elements together:
\begin{equation}
\langle \hat{x}(s) \rangle \sim \left\{
\begin{matrix} s^{-1-p},\ \text{for}\ \gamma_R<1\\
{}\\
 s^{-1},\ \text{for}\ \gamma_R=1\\
\end{matrix} \right. .
\end{equation}
Finally, applying the inverse Laplace transform one finds
\begin{equation}
\langle x(t) \rangle \sim \left\{
\begin{matrix} t^p,\ \text{for}\ \gamma_R<1\\
{}\\
 t^{0},\ \text{for}\ \gamma_R=1\\
\end{matrix} \right.
\end{equation}
\subsection*{A2. Survival probability of a free process with resets}
We proceed similarly to the MSD case. Here we start from Eq. \eqref{Eq14} and we rewrite it as
\begin{equation}
\hat{\sigma}_x(s)=\frac{T_1'(s)}{1-T_2'(s)}
\label{EqsurvApp}
\end{equation}
with $T_1'(s)=\mathcal{L}[\varphi_R^*(t) Q_x(t)]$ and $T_2'(s)=\mathcal{L}[\varphi_R(t) Q_x(t)]$. Let us start for the latter. As in the previous case, we study the small $s$ limit, where we have that
\begin{equation}
T_2'(0)=\int_0^\infty \varphi_R(t) Q_x(t) dt<\int_0^\infty \varphi_R(t) dt=1.
\end{equation}
In the second step we have used that the survival probability $Q_x(t)<1,\ \forall t>0$, and in the last step the normalization of $\varphi_R(t)$ is used. Then, the denominator of Eq. \eqref{EqsurvApp} is strictly positive when $s\rightarrow 0$, which implies that the decaying of $\hat{\sigma}_x(s)$ when $s\rightarrow 0$ is exclusively determined by the decaying of the numerator $T_1'(s)$, i.e.
\begin{equation}
\hat{\sigma}_x(s)\sim \frac{T_1'(s)}{1-T_2'(0)}\sim T_1'(s)
\end{equation}
for small $s$. Applying the inverse Laplace transform to this expression one gets the equivalent relation
\begin{equation}
\sigma_x(t)\sim \mathcal{L}^{-1}[T_1'(s)]=\varphi_R^*(t)Q_x(t)
\end{equation}\\
for long $t$. If the survival probability of the motion process decays as $Q_x(t)\sim t^{-q},\ q>0$, as assumed in the main text, and $\varphi_R^*(t)$ is again the Mittag-Leffler survival probability, which decays as in Eq. \eqref{MLsurvDecay}, we have that
\begin{equation}
\sigma_x(t)\sim t^{-\gamma_R-q}
\end{equation}
asymptotically. On one hand, if $\gamma_R+q\leq 1$ the mean first arrival time is infinite. On the other hand, if $\gamma_R+q>1$, which includes exponentially distributed reset times for $\gamma_R=1$, the mean first arrival time is finite and can be found as
\begin{equation}
T_{F}(x)=\hat{\sigma}_x(0)=\frac{\int_0^\infty \varphi_R^*(t) Q_x(t)dt}{1-\int_0^\infty \varphi_R(t) Q_x(t) dt}
\end{equation}
Finally, taking a Mittag-Leffler reset time distribution, one recovers the result in Eq. \eqref{Eq15} of the main text.
\subsection*{A3. MFAT for exponential motion survival probability}
Here we show that when the motion survival probability is of the form $Q_x(t)= e^{-r(x) t}$, the overall process MFAT is always finite. In this particular case and from Eq. \eqref{Eq13}, the overall survival probability in the Laplace space can be written as
\begin{equation}
\hat{\sigma}_x(s)=\frac{\mathcal{L}[\varphi_R^*(t) e^{-r(x)t}]}{1-\mathcal{L}[\varphi_R(t) e^{-r(x)t}]}=\frac{ \hat{\varphi}_R^*(s+r(x))}{1-\hat{\varphi}_R(s+r(x))}.
\end{equation}
Taking the limit $s\rightarrow 0$ one can get the MFAT:
\begin{equation}
T_F=\lim_{s\rightarrow 0}\hat{\sigma}_x(s)=\frac{\hat{\varphi}_R^*(r(x))}{1- \hat{\varphi}_R(r(x))}
\end{equation}
The Laplace transform of the survival probability can be expressed as
\[
\hat{\varphi}_R^*(s)=\mathcal{L}\left[\int_t^{\infty}\varphi_R(t')dt'\right]=\frac{1-\hat{\varphi}_R(s)}{s},
\]
thus,
\begin{equation}
T_F=\frac{1}{r(x)},
\end{equation}
which is of course finite for all $r(x)$ and independent of $\varphi_R(t)$. In fact, this result is obvious because the completion rate of a process with $Q_x(t)=e^{-r(x) t}$ is constant in time and, therefore, resetting it does not modify its completion time. Regarding the motion survival probability of the process in Section \ref{SecRWPot}, since the finiteness of the MFAT is determined by the long $t$ behavior of the survival probability, when only the asymptotic decay is exponential, the mean first arrival time is also finite.
\bibliography{References}
\end{document}